\documentclass[aps,prc,twocolumn,superscriptaddress,floatfix]{revtex4-1}
\usepackage{textcomp}
\usepackage{nicefrac}
\usepackage{longtable}
\usepackage{graphicx}
\usepackage{multirow}
\usepackage{amsmath}
\usepackage{pbox}
\usepackage{makecell}

\begin{document}
 
 \title{A high-resolution study of levels in the astrophysically important nucleus $^{26}$Mg and resulting updated level assignments}
 
 \author{P. Adsley}
 \email{padsley@gmail.com}
 \affiliation{Institut de Physique Nucl\'{e}aire d'Orsay, UMR8608, CNRS-IN2P3, Universit\'{e} Paris Sud 11, 91406 Orsay, France}
 \author{J.W. Br\"{u}mmer}
 \affiliation{iThemba Laboratory for Accelerator Based Sciences, Somerset West 7129, South Africa}
  \affiliation{Department of Physics,  University of Stellenbosch, Private Bag X1, 7602 Matieland, Stellenbosch, South Africa}
  \author{T Faestermann}
  \affiliation{Physik Department E12, Technische Universit\"{a}t M\"{u}nchen, D-85748 Garching, Germany}
 \author{S. P. Fox}
 \affiliation{Department of Physics, University of York, Heslington, York, YO10 5DD, United Kingdom}
 \author{F. Hammache}
 \affiliation{Institut de Physique Nucl\'{e}aire d'Orsay, UMR8608, CNRS-IN2P3, Universit\'{e} Paris Sud 11, 91406 Orsay, France}
 \author{R. Hertenberger}
 \affiliation{Fakult\"{a}t f\"{u}r Physik, Ludwig-Maximilians-Universit\"{a}t M\"{u}nchen, D-85748 Garching, Germany}
 \author{A. Meyer}
 \affiliation{Institut de Physique Nucl\'{e}aire d'Orsay, UMR8608, CNRS-IN2P3, Universit\'{e} Paris Sud 11, 91406 Orsay, France}
 \author{R. Neveling}
 \affiliation{iThemba Laboratory for Accelerator Based Sciences, Somerset West 7129, South Africa}
 \author{D. Seiler}
 \affiliation{Physik Department E12, Technische Universit\"{a}t M\"{u}nchen, D-85748 Garching, Germany}
 \author{N. de S\'{e}r\'{e}ville}
 \affiliation{Institut de Physique Nucl\'{e}aire d'Orsay, UMR8608, CNRS-IN2P3, Universit\'{e} Paris Sud 11, 91406 Orsay, France} 
 \author{H.-F. Wirth}
 \affiliation{Fakult\"{a}t f\"{u}r Physik, Ludwig-Maximilians-Universit\"{a}t M\"{u}nchen, D-85748 Garching, Germany}
 \date{\today}

\begin{abstract}

\begin{description}
\item[Background] The $^{22}$Ne($\alpha,n$)$^{25}$Mg reaction is an important source of neutrons for the $s$-process. Direct measurement of this reaction and the competing $^{22}$Ne($\alpha,\gamma$)$^{26}$Mg reaction are challenging due to the gaseous nature of both reactants, the low cross section and the experimental challenges of detecting neutrons and high-energy $\gamma$ rays. Detailed knowledge of the resonance properties enables the rates to be constrained for $s$-process models.
\item[Purpose] Previous experimental studies have demonstrated a lack of agreement in both the number and excitation energy of levels in $^{26}$Mg. In order to try to resolve the disagreement between different experiments, proton and deuteron inelastic scattering from $^{26}$Mg have been used to identify excited states.
\item[Method] Proton and deuteron beams from the tandem accelerator at the Maier-Leibnitz Laboratorium at Garching, Munich were incident upon enriched $^{26}$MgO targets. Scattered particles were momentum-analysed in the Q3D magnetic spectrograph and detected at the focal plane.
\item[Results] Reassignments of states around $E_x = 10.8-10.83$ MeV in $^{26}$Mg suggested in previous works have been confirmed. In addition, new states in $^{26}$Mg have been observed, two below and two above the neutron threshold. Up to six additional states above the neutron threshold may have been observed compared to experimental studies of neutron reactions on $^{25}$Mg but some or all of these states may be due to $^{24}$Mg contamination in the target. Finally, inconsistencies between measured resonance strengths and some previously accepted $J^\pi$ assignments of excited $^{26}$Mg states have been noted.
\item[Conclusion] There are still a large number of nuclear properties in $^{26}$Mg which have yet to be determined and levels which are, at present, not included in calculations of the reaction rates. In addition, some inconsistencies between existing nuclear data exist which must be resolved in order for the reaction rates to be properly calculated.
\end{description}

\end{abstract}

\maketitle

\section{Astrophysical Background and Summary of Previous Experimental Studies}

The slow neutron-capture process ($s$-process) is one of the nucleosynthetic processes responsible for the production of elements heavier than iron \cite{RevModPhys.83.157}. The neutrons which contribute to the $s$-process result mainly from two reactions: $^{13}$C($\alpha,n$)$^{16}$O and $^{22}$Ne($\alpha,n$)$^{25}$Mg. The $^{13}$C($\alpha,n$)$^{16}$O reaction is active in thermally pulsing low-mass asymptotic giant branch stars. The $^{22}$Ne($\alpha,n$)$^{25}$Mg reaction is active during thermal pulses in low- and intermediate-mass asymptotic giant branch (AGB) stars and in the helium-burning and carbon-shell burning stages in massive stars (see Ref. \cite{RevModPhys.83.157} and references therein). The $^{22}$Ne($\alpha,n$)$^{25}$Mg reaction is slightly endothermic ($Q = -478.29$ keV, $S_n = 11.093$ MeV) and does not strongly operate until slightly higher temperatures are reached during either the thermal pulse in AGB stars or, in massive stars, at the end of helium burning ($0.25-0.3$ GK, Gamow window: $E_x = 11.025 - 11.365$ MeV). In contrast, the $^{22}$Ne($\alpha,\gamma$)$^{26}$Mg reaction ($S_\alpha = 10.615$ MeV) is able to operate continuously at lower temperatures ($0.1-0.2$ GK), consuming some of the $^{22}$Ne which may otherwise contribute to the total neutron production. Past studies have emphasised the importance of having a complete knowledge of the $^{22}$Ne($\alpha,n$)$^{25}$Mg and $^{22}$Ne($\alpha,\gamma$)$^{26}$Mg reaction rates at a range of temperatures \cite{PhysRevC.85.065809}.

Direct measurements of $^{22}$Ne$+\alpha$ reactions are difficult not only due to the low cross sections involved but also the gaseous nature of both of the species, and the difficulty of detecting neutrons and high-energy $\gamma$ rays. Despite these difficulties, direct measurements of the $^{22}$Ne($\alpha,n$)$^{25}$Mg reaction down to $E_{\alpha}^{\mathrm{lab}} = 570$ keV exist \cite{PhysRevLett.87.202501,1993ApJ...414..735D} along with a simultaneous measurement of the $^{22}$Ne($\alpha,n$)$^{25}$Mg and $^{22}$Ne($\alpha,\gamma$)$^{26}$Mg reactions \cite{Wolke1989}.

In the absence of existing direct measurements at lower temperatures, the knowledge of the properties of resonances in $^{26}$Mg may be used to better-constrain the $^{22}$Ne$+\alpha$ reaction rates. To this end, a number of experimental studies have been performed to probe the properties of levels in $^{26}$Mg. A brief summary of these experimental studies is given so that comparisons to the states observed in the present experiment may be made later.

The $^{26}$Mg($p,p^\prime$)$^{26}$Mg reaction has been measured at a low proton energy \cite{Moss1976429}. The reaction mechanism for this reaction is not selective \cite{PhysRevC.89.065805,413}. Thus, experiments of the type described in Refs. \cite{Moss1976429,PhysRevC.89.065805} may be used as a reference for other experimental works as to how many states are present and the excitation energies of the states.

The $^{26}$Mg($p,p^\prime$)$^{26}$Mg reaction has also been measured at a higher proton energy ($E_p^{\mathrm{lab}} = 200$ MeV) for the purpose of determining the $M1$ strength distribution in $^{26}$Mg \cite{PhysRevC.39.311}. This experiment may be used to identify known $1^+$ states which, being of unnatural parity, cannot contribute to the astrophysical $^{22}$Ne$+\alpha$ reaction rates, for the purposes of excluding said states from the rate calculation.

The $^{26}$Mg($\alpha,\alpha^\prime$)$^{26}$Mg reaction using $E_\alpha = 200$ MeV has been performed twice on roughly comparable experimental setups \cite{PhysRevC.93.055803,PhysRevC.96.055802}. Ref. \cite{PhysRevC.96.055802} suggests that other states which may not have previously been observed may exist in $^{26}$Mg, in particular that there is a previously unresolved multiplet at around $E_{x} = 10.81$ MeV based on the differential cross sections observed combined with data from other experiments. Alpha-particle inelastic scattering is highly selective to isoscalar states with natural parity, i.e. those states which may strongly contribute to the $^{22}$Ne$+\alpha$ reactions. However, the energy resolution of these experiments is insufficient to resolve some of the states observed by Moss \cite{Moss1976429}. Rather, the discernment that additional states are present comes from the differential cross sections and comparisons to other experimental studies of $^{26}$Mg. 

The $^{22}$Ne($^6$Li,$d$)$^{26}$Mg reaction has been measured at a number of different beam energies \cite{PhysRevC.93.055803,PhysRevC.76.025802,GIESEN199395,refId0}. This reaction should preferentially populate natural-parity isoscalar states with large $\alpha$-particle reduced widths, i.e. states with an $\alpha$-particle cluster structure. From the comparison of the cross section of these reactions with DWBA calculations, it is possible to extract the $\alpha$-particle spectroscopic factor and then to calculate the $\alpha$-particle partial widths of the states albeit with large uncertainties due to the modelling of the reaction mechanism. Previous studies of the $^{22}$Ne($^6$Li,$d$)$^{26}$Mg reaction have had quite poor energy resolution, $120$ keV in Ref. \cite{GIESEN199395}, $60-70$ keV for Ref. \cite{PhysRevC.76.025802} and 100 keV for Ref. \cite{PhysRevC.93.055803}. It is possible that some of the states observed in these reactions may in fact be multiple states in close proximity resulting in differential cross sections that consist of multiple contributions thus making extraction of the $\ell$-value and spectroscopic factors from this reaction difficult to interpret.

The $^{26}$Mg($\gamma,\gamma^\prime$)$^{26}$Mg reaction has been measured using polarised $\gamma$ rays at the HI$\gamma$S facility \cite{PhysRevC.80.055803,PhysRevC.82.025802} and unpolarised $\gamma$ rays at ELBE \cite{PhysRevC.79.037303}. These studies allow the $\gamma$-ray partial widths to be determined and $J^\pi$ assignments to be made. However, $\gamma$-ray inelastic scattering is primarily limited to the observation of low-spin states, and states with $J=0$ cannot be directly observed.

Finally, the $^{25}$Mg($n,\gamma$)$^{26}$Mg radiative capture and $^{25}$Mg($n$,tot) transmission reactions have been measured \cite{PhysRevC.85.044615,Massimi20171,PhysRevC.66.055805}. These reactions are primarily sensitive to states above the neutron threshold and so are unable to clarify, for example, the discrepancies which are suggested in Ref. \cite{PhysRevC.96.055802}. In addition, the nature of the neutron-induced reaction means that states which have small neutron widths will not be observed in either the radiative capture or transmission measurements. This leaves open the possibility that $^{25}$Mg$+n$ experiments may miss states with inhibited neutron decay channels. It is important to verify that no levels have been missed by this neutron-induced study to avoid potential bias in the calculation of the reaction rates.

To attempt to resolve the discrepancies between Refs. \cite{Moss1976429}, \cite{PhysRevC.93.055803}, and \cite{PhysRevC.96.055802} on the location and $J^\pi$ assignments of the excited states in $^{26}$Mg, and to investigate if other levels in $^{26}$Mg were not located in Ref. \cite{Moss1976429} we have repeated the $^{26}$Mg($p,p^\prime$)$^{26}$Mg measurement of Moss \cite{Moss1976429} using the Q3D magnetic spectrograph at the Maier-Leibnitz Laboratorium, Munich.

In addition to this measurement, another experiment using the $^{26}$Mg($d,d^\prime$)$^{26}$Mg reaction was also performed. Performing deuteron scattering in addition to proton scattering provides two benefits. Firstly, the kinematics of deuteron scattering are significantly different to proton scattering due to the differing ratio of projectile mass to target mass. This means that contaminant states on the focal plane shift significantly between the proton and deuteron scattering data giving an additional verification for levels in $^{26}$Mg. Secondly, the inelastic scattering of deuterons has selectivity to isoscalar transitions \cite{Kawabata20076}. As $^{22}$Ne has isospin $T=1$ and the $\alpha$ particle has $T=0$, the states in $^{26}$Mg which can contribute to the $^{22}$Ne$+\alpha$ reactions must also have $T=1$. The inelastic scattering of the deuteron, which is also $T=0$, should preferentially populate $T=1$ states in $^{26}$Mg, the ground state of which has $T=1$. This can provide valuable information as to which observed states are able to contribute to the $^{22}$Ne$+\alpha$ reactions; states which are not populated in ($d,d^\prime$) reactions likely have small $\Gamma_\alpha$ widths and contribute weakly to the $^{22}$Ne$+\alpha$ reactions.

\section{Experimental Method}

Proton and deuteron beams ($E_{\mathrm{beam}} = 18$ MeV) from the tandem accelerator at the Maier-Leibnitz Laboratorium were incident upon a target consisting of 40-$\mu$g/cm$^2$ of $^{26}$MgO (enrichment of $^{26}$Mg: 94\% determined by elastic scattering of deuterons at 40 degrees) on a 20-$\mu$g/cm$^2$ $^{12}$C backing. Reaction products were momentum-analysed in the MLL Q3D magnetic spectrograph \cite{LOFFLER19731}. Focal-plane particle identification was achieved considering the energy deposited in the two gas detectors and a plastic scintillator at the focal plane of the spectrograph. 

In addition to the data taken with the $^{26}$MgO target, background data were taken with a carbon target identical to that used for the target backing; a flat background was observed from the carbon data. Data were also taken with a silicon oxide target for the purposes of calibrating the focal plane and characterisation of the oxygen background.

Proton- and deuteron-scattering data were taken with the field setting covering from around $E_x = 10.6-11.1$ MeV in $^{26}$Mg at 35 and 40 degree scattering angles. By collecting data at multiple angles, it is possible to identify peaks on the focal plane resulting from target contaminants; peaks resulting from target contaminants shift on the focal plane relative to states in the target of interest when changing angle. Proton-scattering data were also taken at $E_x = 10.9-11.5$ MeV at 40 degrees only.

\section{Data Analysis}

Scattered protons or deuterons were selected at the focal plane of the Q3D using software gates on the energy deposited in the proportional counters and the plastic scintillator. The focal plane was calibrated in magnetic rigidity, $B\rho$, using well-known isolated states in $^{28}$Si and taking into account the energy loss of the scattered proton or deuteron in the target. The calibration data were taken using the magnetic field settings for the $^{26}$Mg data. From $B\rho$, the detected proton or deuteron energy was calculated, corrected for energy losses in the target and then used to calculate the corresponding excitation energy in $^{26}$Mg. Energy losses in carbon, silicon oxide and magnesium oxide were all taken from the programme \textsc{dedx} \cite{dedx}. This procedure is validated by ensuring that the excitation energies of the 10.806- and 10.949-MeV levels observed in $^{26}$Mg($\gamma,\gamma^\prime$)$^{26}$Mg reactions \cite{PhysRevC.80.055803} are reproduced correctly. The experimental resolution for the proton (deuteron) scattering data was 6 (8) keV, FWHM.

Spectra are fitted with a combination of Gaussian peaks for narrow states (those with widths less than the experimental resolution) and Voigt functions for broader states. All of the states in a spectrum use a common experimental resolution. In the spectra resulting from proton scattering, the $^{16}$O states are described by exponentially-tailed Gaussian functions given by \cite{das2016simple}:

 $f(x; \mu, \sigma, \kappa) = 
 \begin{cases}
  A e^{(x - \mu)^2 / 2\sigma^2}  &\kappa \geq \frac{x-\mu}{\sigma} \\
  A e^{\kappa^2/2 - \kappa (x-\mu)/\sigma)} &\kappa < \frac{x-\mu}{\sigma} \\
 \end{cases}
$

\noindent where $A$ is the amplitude of the functions, $\mu$ is centroid energy, $\sigma$ the resolution for the contaminant state (which differs from the common experimental resolution used for the $^{26}$Mg states) and $\kappa$ is the matching parameter giving the number of standard deviations from the centroid where the function switches from the Gaussian form to the exponential form. All states below the neutron threshold and any state above the neutron threshold which did not appear in the $^{25}$Mg$+n$ data of Refs. \cite{PhysRevC.85.044615} and \cite{Massimi20171} is assumed to be narrow; these states are fitted with Gaussian functions. This is because, for $^{26}$Mg states in the excitation-energy region being investigated, the width for a broad state must be dominated by the neutron width and the $^{25}$Mg$+n$ reactions are sensitive to any state with a neutron width above around $0.5$ eV (see the discussion in Ref. \cite{Massimi20171} for details).

In the deuteron-scattering spectra, the region containing the $^{16}$O 10.356-MeV contaminant state is omitted from the fit but the contribution of this state to the spectrum was accounted for using a Gaussian function for which the centroid and variance parameters were determined from the silicon oxide calibration target.

All spectra include an additional quadratic polynomial background which accounts for the various other sources of background such as multiple scattering within the spectrograph, continuum effects and broad states in, for example, the carbon from the target backing.

\begin{figure*}
\includegraphics[width=\textwidth]{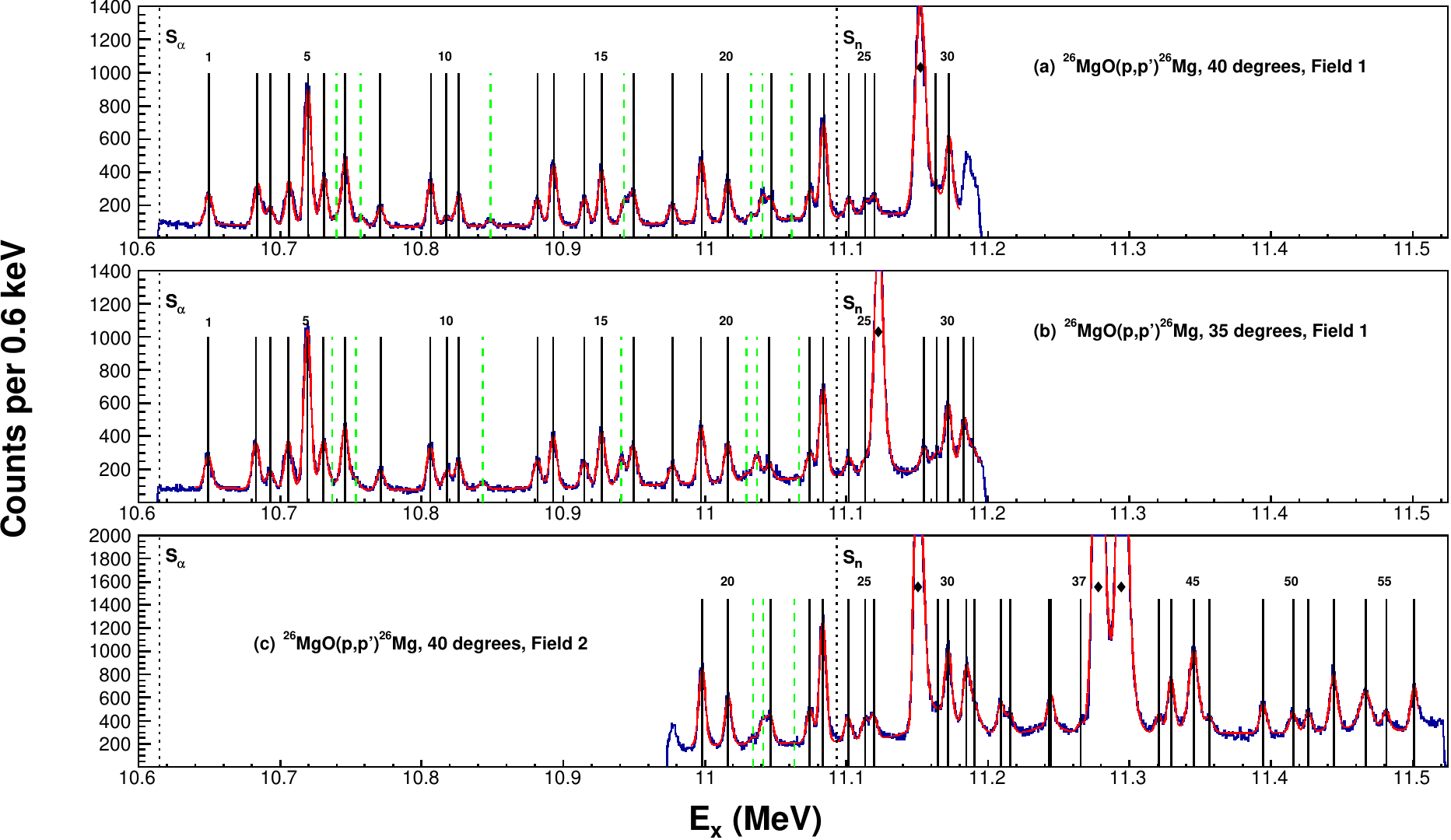}
 \caption{Excitation-energy spectra for $^{26}$Mg. See the figure for details of each spectrum. Vertical black lines denote a state which is observed at multiple angles, green dashed lines denote a contaminant peak. Black diamonds mark the $^{16}$O contaminant peaks. The solid red line is the fit.}
 \label{fig:protonspectra}
\end{figure*}

\begin{figure*}
 \includegraphics[width=\textwidth]{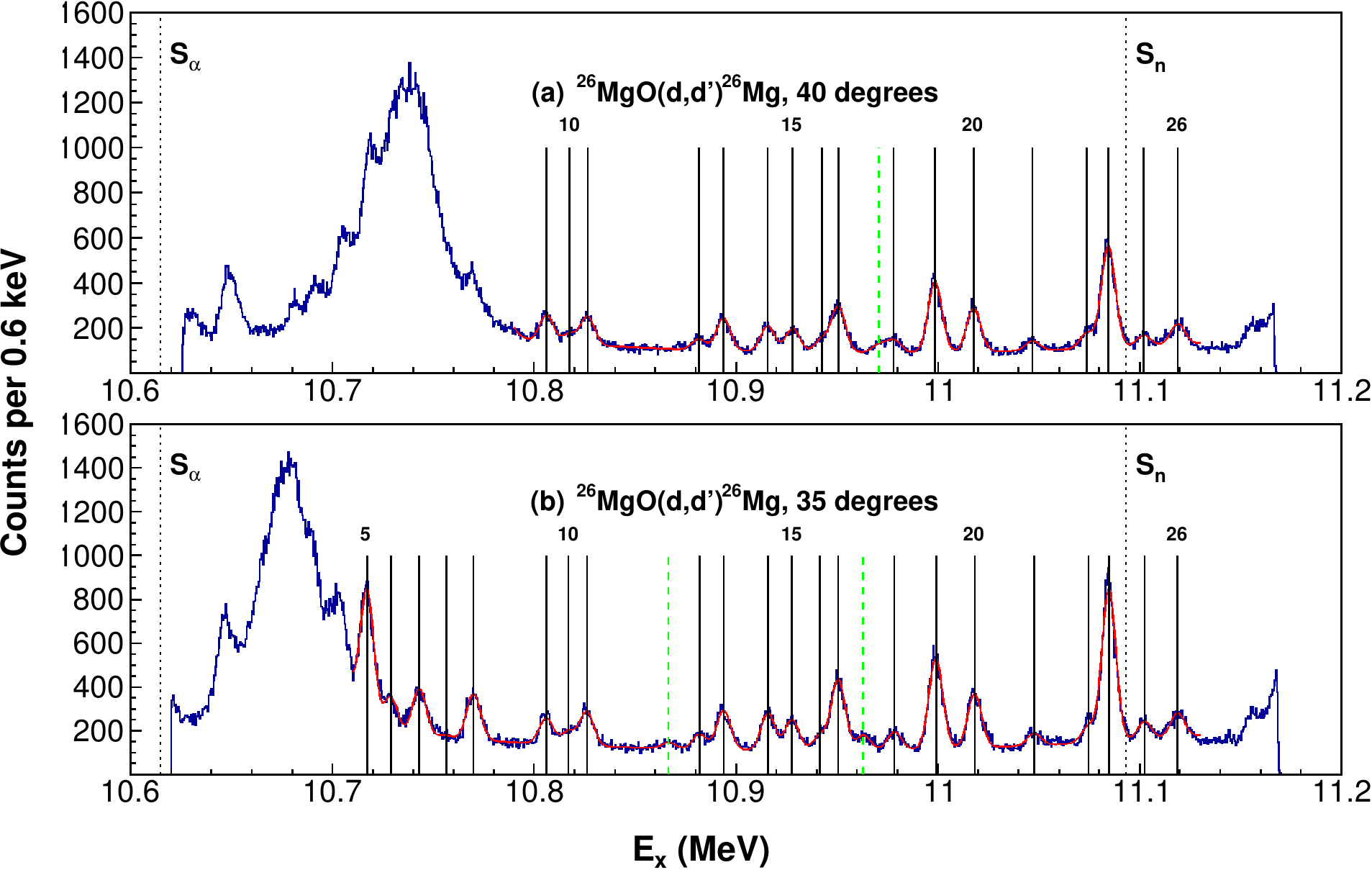}
 \caption{Excitation-energy spectra for $^{26}$Mg. See the figure for details of each spectrum. Vertical black lines denote a state which is observed at multiple angles, green dashed lines denote a contaminant peak. The solid red line is the fit.}
 \label{fig:deuteronspectra}
\end{figure*}

The obtained excitation-energy spectra are shown in Figures \ref{fig:protonspectra} and \ref{fig:deuteronspectra}.

\section{Results and Discussion}

A summary of the levels observed in this experiment are given in Table \ref{tab:states} along with suggested correspondences with levels observed in other experimental studies, and resulting spin and parity assignments. For details as to the assignments made, see the text and Table \ref{tab:states}. Only states where the assignment is unclear or inconsistent with other nuclear data, or generally in need to clarifying remarks are discussed in the text. The discussion of the assignments is split into two sections, one below the neutron threshold for which comparison to the $^{25}$Mg$+n$ data of Refs. \cite{Massimi20171} and \cite{PhysRevC.85.044615} does not need to be made, and the other above the neutron threshold. Each of the states is given an index number in the first column of Table \ref{tab:states} for ease of reference. These state indices are used both in the discussion of the level assignments and also in Figures \ref{fig:protonspectra} and \ref{fig:deuteronspectra}. Note that some of the state indices refer to levels observed in other experiments but not in the present experiments (due to, for example, contaminating $^{16}$O levels) and that these states do not appear in the spectra in Figs. \ref{fig:protonspectra} and \ref{fig:deuteronspectra}.

\LTcapwidth=\textwidth
\begin{longtable*}[t]{c|c|c|c|c|c|c|c}
\newpage
\caption{Excitation energies of $^{26}$Mg states observed in the present study with suggested $J^\pi$ assignments and comparisons to previous experimental measurements. See the text for explanations of the assignments made for the states. The errors given in the table for the present experiment are statistical only. For a discussion about the sources of systematic error, see the text. The errors for Refs. \cite{Massimi20171,PhysRevC.85.044615} are omitted as all are much smaller than 1 keV.}\\ \hline
\label{tab:states}
  Index & \thead{$E_x$\\(MeV)\\This paper \footnote{Error listed for states is statistical only. See the text for a discussion as to the systematic error.}} & \thead{Recommended\\ $J^\pi$} & \thead{$E_x$ (MeV)\\$^{26}$Mg($\gamma,\gamma^\prime$)\\ \cite{PhysRevC.80.055803}}  & \thead{$E_x$ (MeV)\\$^{26}$Mg($p,p^\prime$)\\ \cite{Moss1976429}\\} & \thead{$E_x$ (MeV)\\$^{25}$Mg$+n$\\ \cite{Massimi20171,PhysRevC.85.044615}} & \thead{$E_x$ (MeV)\\$^{22}$Ne($\alpha,n$) \\\cite{PhysRevLett.87.202501}} & Comments \\
  \hline \hline
   1 & 10.650(1) & $1^+$ & 10647.3(8) & 10.644(3) & & & $J^\pi$ from Ref. \cite{PhysRevC.80.055803}.  \\
   2 & 10.684(2) & & & 10.678(3) & & & \\
   3 & 10.693(1) & $4^+$ & & 10.689(3) & & & \\
   4 & 10.706(1) & &  & 10.702(3) & & & \\
   5 & 10.719(2) & $2^+$ & & 10.715(3) & & & \\
   6 & 10.730(2) & & & 10.726(3) & & & \\
   7 & 10.746(3) & & & 10.744(3) & & & \\
   8 & 10.771(1) & & & 10.769(3) & & & \\
   9 & 10.806(1) & $1^-$ & 10805.7(7)  &   &        & & $J^\pi$ from Ref. \cite{PhysRevC.80.055803}. \\
    & & & & & & & \\
   10 & 10.818(1) & $1^+$ &  &   &  & & \makecell{Assumed to be the state at\\ $E_x = 10.81$ MeV from Ref. \cite{PhysRevC.39.311}.} \\
   & & & & & & & \\
   11 & 10.826(1) & $0^+$ & & 10.824(3) & & & \makecell{Assumed to be the state at\\ $E_x = 10.82$ MeV from Ref. \cite{PhysRevC.96.055802}.} \\
    & & & & & & & \\
   12 & 10.882(1) & & & 10.881(3) & & & \\
   13 & 10.893(1) & & & 10.893(3) & & & \\
   14 & 10.915(1) & & & 10.915(3) & & & \\
   15 & 10.928(1) & & & 10.927(3) & & & \\
    & & & & & & & \\
   16 & 10.943(2) & & & & & & \makecell{Possible new state, seen only in \\ $^{26}$Mg($d,d^\prime$)$^{26}$Mg. See text for details.}\\
    & & & & & & & \\
   17 & 10.950(1) & $1^-$ & 10949.1(8) & 10.950(3) & & & $J^\pi$ from Ref. \cite{PhysRevC.80.055803}. \\
   18 & 10.978(1) & & & 10.978(3) & & & \\
   19 & 10.998(1) & & & 10.998(3) & & & \\
   20 & 11.017(1) & & & 11.017(3) & & & \\
   21 & 11.047(1) & & & 11.048(3) & & & \\
   22 & 11.074(1) & & & & & & New state \\
   23 & 11.084(1) & & & 11.084(3) & & & \\
   24 & 11.102(1) & & & & & & New state \\
   25 & 11.113(1) & $2^+$ & & 	   & 11.112 & & Not seen in $^{26}$Mg($d,d^\prime$), possibly $T=2$ \\
   26 & 11.119(1) & & & & & & New state\\
    & & & & & & & \\
   27 & 11.155(1) & $1^+$ & 11153.5(10) & 11.156(3) & 11.154 & &  \makecell{Only observed at one angle\\$J^\pi$ from Ref. \cite{PhysRevC.80.055803}.} \\
    & & & & & & & \\
   28 & 11.165(1) & $2^+$ & &           & 11.163 & & \makecell{$J^\pi$ from Ref. \cite{Massimi20171}. \\ See note in the text about this level.} \\
    & & & & & & & \\
   29 & 11.165(1) & $3^-$ & &           & 11.169 & & \makecell{$J^\pi$ from Ref. \cite{Massimi20171}. \\ See note in the text about this level.} \\
    & & & & & & & \\
   30 & 11.172(1) & & & 11.171(3) & 11.171 & & \\
   31 & 11.184(1) & $(1^-)$ & & & 11.183 & & $J^\pi$ from Ref. \cite{PhysRevC.85.044615}.\\
   32 & 11.191(1) & $3^+$ & &           & 11.190 & & $J^\pi$ from Ref. \cite{Massimi20171}.\\
    & & & & & & & \\
   33 & 11.209(1) & & & & & & \makecell{Only at one angle.\\Possible $^{24}$Mg contaminant,\\$E_{x,^{24}\mathrm{Mg}} = 11.181$ MeV.} \\
    & & & & & & & \\
   34 & 11.216(1) & & & & & & \makecell{Only at one angle.\\Possible $^{24}$Mg contaminant,\\$E_{x,^{24}\mathrm{Mg}} = 11.186$ MeV.} \\
    & & & & & & & \\
   35 & 11.243(3) & & & & & & $\Gamma = 29(3)$ keV. See text. \\
   36 & 11.245(1) & $2^-$ & &          & 11.243 & & $J^\pi$ from Ref. \cite{Massimi20171}. See text. \\
   37 & 11.266(1) & & & & & & \makecell{Only at one angle.\\Possible new state.} \\
   38 & & $2^+$ & &           & 11.274 & & \makecell{Obscured by contaminant.\\Data from Ref. \cite{Massimi20171}.} \\
    & & & & & & & \\
   39 & & $3^-$ & &           & 11.280 & & \makecell{Obscured by contaminant.\\Data from Ref. \cite{Massimi20171}.} \\
    & & & & & & & \\
   40 & & $2^-$ & &           & 11.285 & & \makecell{Obscured by contaminant.\\Data from Ref. \cite{Massimi20171}.} \\
    & & & & & & & \\
   41 & & $>1$ & &           & 11.289 & & \makecell{Obscured by contaminant.\\$E_x$ from Ref. \cite{Massimi20171}, $J^\pi$ from Ref. \cite{PhysRevC.96.055802}.,\\Natural parity.} \\
    & & & & & & & \\
   42 & & $2^-$ & &           & 11.295 & & \makecell{Obscured by contaminant.\\ Data from Ref. \cite{Massimi20171}.} \\
   43 & 11.321(1) & & & & & 11.319(2) & \\
    & & & & & & & \\
   44 & 11.329(1) & $(1^+)$ & &           & 11.328 & & \makecell{$J^\pi$ from Ref. \cite{PhysRevC.39.311}. See text for details.}  \\
    & & & & & & & \\
   45 & 11.345(1) & & &       &    & 11.344 & \makecell{Two states in Ref. \cite{Massimi20171}.\\See text for details.} \\
    & & & & & & & \\
   46 & & $>3$ & &           & 11.344 & & \makecell{$J^\pi$ from Ref. \cite{Massimi20171}.\\See note for state above and the text.} \\
    & & & & & & & \\
   47 & 11.357(1) & & & & & \\
   48 & & & & & 11.362 & & Not observed in the present experiment. \\
   49 & 11.395(1) & & & & 11.393 &  &\\
   50 & 11.414(1) & & & & & & \makecell{Only at one angle.\\Possible $^{24}$Mg contaminant,\\$E_{x,^{24}\mathrm{Mg}} = 11.389$ MeV.} \\
   & & & & & & & \\
   51 & 11.426(1) & & & & & & \makecell{Possible new or $^{24}$Mg contaminant state.\\$E_{x,^{24}\mathrm{Mg}} = 11.453$ MeV.} \\
    & & & & & & & \\
   52 & 11.444(1) & $(4^+)\longrightarrow J\leq3$ & & & 11.441 & 11.441(2) & \makecell{$J^\pi$ assignment from Refs.\\ \cite{PhysRevC.66.055805,Massimi20171,PhysRevC.85.044615} is problematic - see text.} \\
    & & & & & & & \\
   53 & 11.46(1) & $1^+$ & & & &  & \makecell{$J^\pi$ from Ref. \cite{PhysRevC.39.311}. May be the\\ state observed in Refs. \cite{PhysRevC.66.055805,Massimi20171,PhysRevC.85.044615}.} \\
    & & & & & & & \\
   54 & 11.467(1) & $(5^-)\longrightarrow J\leq3$ & & & 11.466 & 11.461(2) & \makecell{$J^\pi$ assignment from Refs.\\ \cite{PhysRevC.66.055805,Massimi20171,PhysRevC.85.044615} is problematic - see text.}  \\
    & & & & & & & \\
   55 & 11.481(1) & & & & & & \makecell{Only at one angle.\\Possible $^{24}$Mg contaminant,\\$E_{x,^{24}\mathrm{Mg}} = 11.456$ MeV.} \\
    & & & & & & & \\
   56 & 11.501(1) & & & & 11.500 & & \\
   \hline \hline
\end{longtable*}

\null\clearpage

The excitation energies of the levels given in Table \ref{tab:states} are taken from the arithmetic weighted mean,

\begin{equation}
 \bar{x} = \frac{1}{\sum_i^N \frac{1}{\sigma_i^2}} \sum_i^N \frac{X_i}{\sigma_i^2}, 
\end{equation}

of the observed levels in all of the spectra in which that state appears. The associated statistical deviation $\sigma_{\bar{x}}$ on the excitation energies,

\begin{equation}
 \sigma_{\bar{x}}^2 = \frac{1}{\sum_i^N \frac{1}{\sigma_i^2}} \frac{1}{N-1} \sum_i^N \frac{(X_i - \bar{x})^2}{\sigma_i^2},
\end{equation}
is also given for each state in Table \ref{tab:states}.

To account for  systematic errors, the variations in excitation energy resulting from various sources of systematic error are computed in Table \ref{tab:uncertainties}. The effect of the beam energy shift on the excitation is small. This is because the beam energy is one of the inputs to the calibration of the focal plane position and is subsumed into that calibration with a minimal effect of the resulting excitation energy calculation.

The uncertainty resulting from shifts in the spectrograph fields or beam energy during the experiment from whatever source was estimated by fitting some of the stronger experimental peaks for subsets of events to look for possible variations. Variations of no more than 0.5 keV were observed and so this was assumed to be the systematic uncertainty resulting from possible field shifts.

\begin{table}
\caption{Potential sources of systematic error and the corresponding contribution to the systematic error.}
\label{tab:uncertainties}
 \begin{tabular}{c|c|c}
 \hline
  Source & \thead{Assumed\\Uncertainty} & \thead{Resulting $E_x$\\uncertainty ($1\sigma$)} \\
  \hline \hline
  Angle & $0.1$ degrees & 1 keV \\
  Target Thickness (MgO) & $10\%$ & 0.1 keV \\
  Target Thickness (C) & $10\%$ & 0.1 keV \\
  Energy Loss & $10\%$ & 0.1 keV\\
  Beam Energy & 2 keV & 0.1 keV \\
  Field shifts & \makecell{Determined\\ from data} & 0.5 keV \\
  \hline
  Total & & 1.1 keV
 \end{tabular}
\end{table}

The total systematic uncertainty is taken as the uncorrelated sum in quadrature of the various components and amounts to $1.1$ keV at the $1\sigma$ level.

The systematic uncertainty of the excitation energies of the states is correlated and, because of this, it is given separately from the statistical uncertainty for each state so that proper account for the correlated uncertainties on the excitation energies may be made in future Monte Carlo calculations of the $^{22}$Ne$+\alpha$ reaction rates in the manner described in Ref. \cite{longland2017correlated}.

\begin{figure}
 \includegraphics[width=0.5\textwidth]{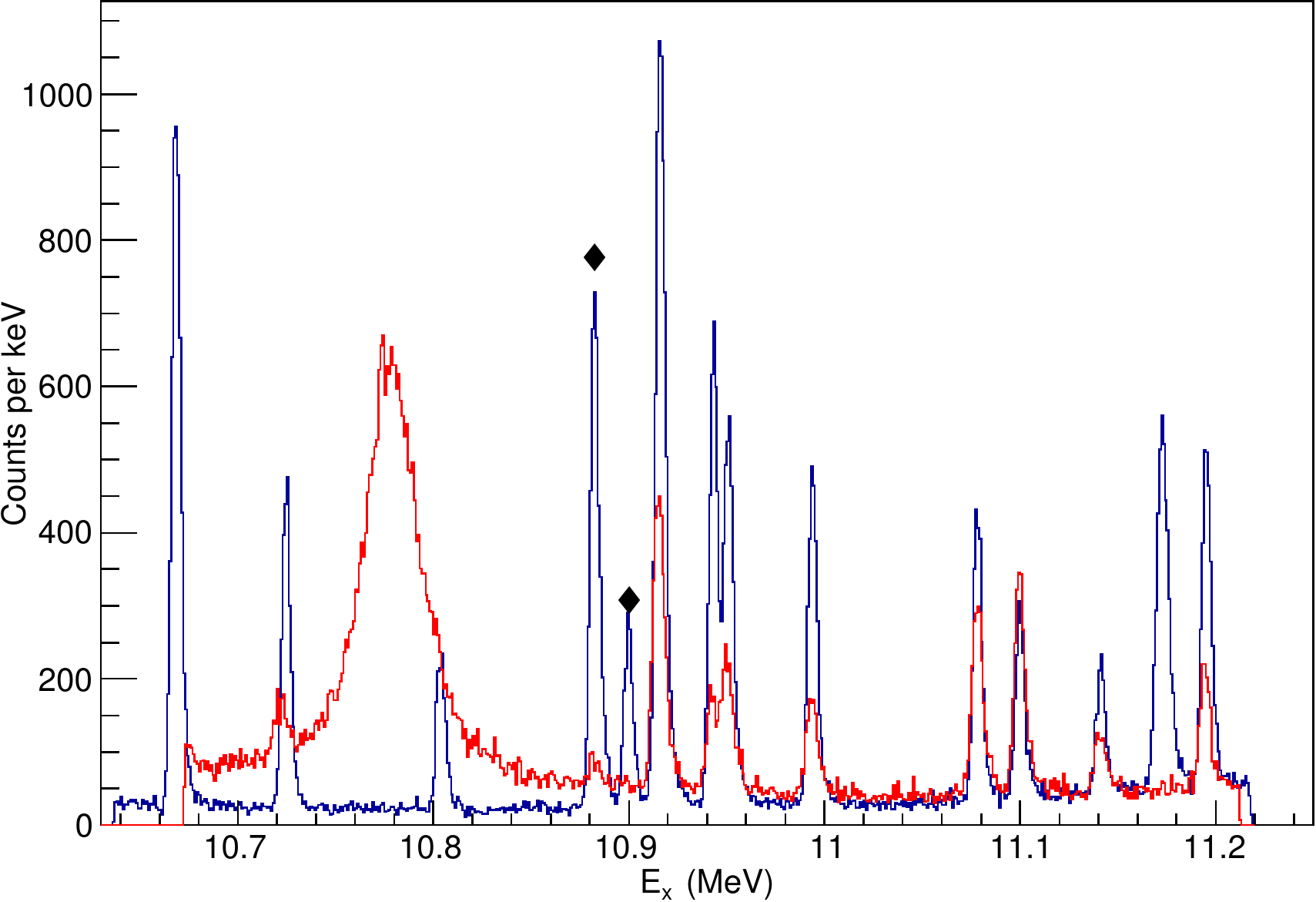}
 \caption{Comparison of $^{28}$Si($p,p^\prime$)$^{28}$Si (blue) and $^{28}$Si($d,d^\prime$)$^{28}$Si (red) spectra at 40 degrees. The suppression of the $T=1$, 10.883- and 10.900-MeV states (diamonds) in $^{28}$Si($d,d'$) is clear. The peak at 11.173 MeV which only appears in the ($p,p^\prime$) spectrum is due to $^{16}$O contamination. The broad state visible between 10.7 and 10.8 MeV in the $^{28}$Si($d,d^\prime$)$^{28}$Si spectrum is due to $^{16}$O contamination.}
 \label{fig:SiliconSpectra}
\end{figure}

To demonstrate the efficacy of the ($d,d^\prime$) reaction in suppressing $\Delta T\neq0$ states we used the $^{28}$Si($p,p^\prime$)$^{28}$Si and $^{28}$Si($d,d'$)$^{28}$Si reactions from the calibration target. Figure \ref{fig:SiliconSpectra} shows the spectra resulting from $^{28}$Si($p,p^\prime$)$^{28}$Si and $^{28}$Si($d,d'$)$^{28}$Si reactions at $\theta_{\mathrm{lab}} = 40$ degrees. The known $T=1$ states at 10.883 and 10.900 MeV \cite{ENSDF} (marked with black diamonds in Figure \ref{fig:SiliconSpectra}) are strongly suppressed in the $^{28}$Si($d,d^\prime$)$^{28}$Si reaction compared to to the $^{28}$Si($p,p^\prime$)$^{28}$Si reaction.
 
\subsection{Between the $\alpha$-particle threshold and the neutron threshold}

\subsubsection{States 9, 10 and 11: The 10.8-10.84 MeV region}

In this region, Moss observed only a single level at 10.824 MeV and connected this level to a $2^+$ level observed in $^{26}$Mg($e,e^\prime$)$^{26}$Mg reactions at 10.838(24) MeV \cite{Moss1976429,0301-0015-7-8-005}. The high-energy $^{26}$Mg($p,p^\prime$)$^{26}$Mg experiment of Crawley {\it et al.} observed a $J^\pi = 1^+$ state at 10.81 MeV \cite{PhysRevC.39.311}. A $\gamma$-ray inelastic-scattering measurement observed a $J^\pi = 1^-$ state at 10.806 MeV \cite{PhysRevC.80.055803}. An $\alpha$-particle inelastic scattering measurement observed a $J^\pi = 0^+$ state at $E_x = 10.82$ MeV \cite{PhysRevC.96.055802} though this disagrees with another $^{26}$Mg($\alpha,\alpha^\prime$)$^{26}$Mg measurement \cite{PhysRevC.93.055803}.

In the present experiment, three states are observed in this region. The energy of the $J^\pi = 1^-$ state is known to be 10.8057(7) MeV \cite{PhysRevC.80.055803} which is in good agreement with the present result of 10.806(1) MeV. The ordering of the other two levels is not definite. The $J^\pi = 1^+$ state of Crawley {\it et al.} was observed at 10.81(1) MeV \footnote{The study of Crawley {\it et al.} gives a resolution of 60 keV but apparently no uncertainty on the excitation energy. However, the comparison between the energies of $1^+$ states in this paper with those observed in Ref. \cite{PhysRevC.79.037303} leads us to conclude that the uncertainty in the excitation energy is around 10 keV).}. The $J^\pi = 0^+$ state observed in $^{26}$Mg($\alpha,\alpha^\prime$)$^{26}$Mg is observed at 10.824(10) MeV in Ref. \cite{PhysRevC.96.055802}. Note that this level energy was fixed in Ref. \cite{PhysRevC.96.055802} according to the energy of the level observed by Moss \cite{Moss1976429}. We take the lower of the two levels to be the $1^+$ state and the higher as the $0^+$ state.

In summary, we conclude that there are three levels in $^{26}$Mg in this region: a $J^\pi = 1^-$ state at 10.806 MeV, a $J^\pi = 1^+$ state at 10.818 MeV and a $J^\pi = 0^+$ state at 10.826 MeV.

Finally, regarding the $2^+$ state observed in the $^{26}$Mg($e,e^\prime$)$^{26}$Mg reaction at $E_x = 10.838(24)$ MeV \cite{0301-0015-7-8-005} which Moss suggested was the single state observed at 10.824 MeV \cite{Moss1976429}: we see no candidate for this state and instead suggest that the observed structure in $^{26}$Mg($e,e^\prime$) may have been a combination of the three states observed in the present experiment rather than a distinct state.

\subsubsection{State 16: 10.943 MeV}

A state is observed in $^{26}$Mg($d,d^\prime$)$^{26}$Mg at $E_x = 10.943(2)$ MeV at both angles. In the $^{26}$Mg($p,p^\prime$)$^{26}$Mg a state is observed at this excitation energy but shifts with angle meaning that it is a contaminant peak. The state observed in $^{26}$Mg($d,d^\prime$)$^{26}$Mg reactions is likely obscured by this contaminant peak in the $^{26}$Mg($p,p^\prime$)$^{26}$Mg data meaning that it is not observed.

\subsubsection{State 22: 11.074 MeV}

This state lies just below the neutron threshold. No information on the spin or parity of this state is available. This state was not observed in the previous high-resolution $^{26}$Mg($p,p^\prime$)$^{26}$Mg experiment \cite{Moss1976429}.

\subsection{Above the neutron threshold}

\subsubsection{State 24: 11.102 MeV}

A new state is observed at $E_x = 11.102$ MeV corresponding to $E_n^{\mathrm{lab}} = 9$ keV in $^{25}$Mg$+n$ experiments. This state is observed in all spectra. This state was not observed in the $^{25}$Mg$+n$ reactions of Refs. \cite{Massimi20171} and \cite{PhysRevC.85.044615} which implies that this state has a small neutron width.

\subsubsection{State 25: 11.113 MeV}

A state is observed in the proton-scattering data at $E_x = 11.113$ MeV ($E_n^{\mathrm{lab}} = 20$ keV). In the 35-degree $^{26}$Mg($p,p^\prime$)$^{26}$Mg data, this state is extremely close to the contaminating state from $^{16}$O and so the assignment is tentative. However, there is a known $J^\pi = 2^+$ state observed in $^{25}$Mg$+n$ measurements at $E_n^{\mathrm{lab}} = 19.86$ keV \cite{Massimi20171,PhysRevC.85.044615}. This state is not observed in the deuteron-scattering data implying that it may not have $T=1$ and thus have a small contribution to the $^{22}$Ne$+\alpha$ reactions.

\subsubsection{State 27: 11.155 MeV}

This state is only observed in the $^{26}$Mg($p,p^\prime$)$^{26}$Mg data at 35 degrees; in the 40-degree data the state is obscured by a contaminating $^{16}$O state. This state corresponds to the known $J^\pi = 1^+$ level observed in $^{25}$Mg$+n$ \cite{Massimi20171} and $^{26}$Mg($\gamma,\gamma^\prime$) \cite{PhysRevC.80.055803} reactions. This $J^\pi = 1^+$ state was also observed at $E_x = 11.15(1)$ MeV in $^{26}$Mg($p,p^\prime$)$^{26}$Mg reactions at $E_p = 200$ MeV \cite{PhysRevC.39.311}. The state is observed close to the end of the focal plane in the $^{26}$Mg($d,d^\prime$)$^{26}$Mg spectra outside the fit region.

\subsubsection{States 28 and 29: 11.165 MeV}

The results of Massimi {\it et al.} \cite{Massimi20171} show states at 11.163 and 11.169 MeV. In the present experiment, only one state is observed at this excitation energy. However, the states may not be resolved in the present experiment.

Note that the two states observed in $^{25}$Mg$+n$ reactions are listed in Table \ref{tab:states} despite only one being observed in the present experiment.

\subsubsection{State 31: 11.184 MeV}

A state is observed at 11.184 MeV in the $^{26}$Mg($p,p^\prime$)$^{26}$Mg data. This state is likely to be the $J^\pi = 1^-$ state which has been observed in one $^{25}$Mg$+n$ experiment \cite{PhysRevC.85.044615} but omitted in another \cite{Massimi20171}. As this state has a narrow neutron width in Ref. \cite{PhysRevC.85.044615}, it is probably below the limit-of-detection for Ref. \cite{Massimi20171}. From Ref. \cite{PhysRevC.85.044615}, this state has a tentative $J^\pi = 1^-$ assignment, meaning that it may contribute to the $^{22}$Ne$+\alpha$ reactions.

\subsubsection{State 32: 11.191 MeV}

In the present data, one state is observed at 11.191 MeV with $\Gamma = 5.2(8)$ keV. We assume that this is the $J^\pi = 3^+$ state observed in Ref. \cite{Massimi20171} which has $\Gamma = 5.24(4)$ keV. We note, however, that Ref. \cite{PhysRevC.85.044615} also includes a tentative state at 11.191 MeV; $J^\pi = 2^-$. No evidence of this tentative state is found in the present experiment.

\subsubsection{States 33 and 34: 11.209 MeV and 11.216 MeV}

Two states are observed at 11.209 and 11.216 MeV corresponding to $E_n^{\mathrm{lab}} = 121$ and $128$ keV respectively. Lacking confirmatory data from a second angle, it is not possible to assign these states definitively to $^{26}$Mg or to reject them as contaminants.


Two states are observed in $^{24}$Mg at $E_x = 11.181$ and $11.186$ MeV which would correspond to $E_x = 11.207$ and $11.212$ MeV in $^{26}$Mg taking into account the kinematic shift. These states could correspond to the states observed in the present experiment.

If these states are real then the neutron widths for both must be small to have escaped detection in previous $^{25}$Mg$+n$ experiments \cite{Massimi20171,PhysRevC.85.044615}.

\subsubsection{States 35 and 36: 11.243 MeV and 11.245 MeV}

Two states are required to fit the spectrum at this energy, a narrow state at 11.245 MeV which likely corresponds to the state observed by Massimi {\it et al.} at 11.243 MeV ($\Gamma = 5950(50)$ eV \cite{Massimi20171}) and a broader state centered on 11.243 MeV with $\Gamma = 29(3)$ keV. There is nothing in the carbon or silicon oxide background spectra that suggest the presence of a contaminant state at this excitation energy. Only having data at one angle we are unable to confirm the existence of a broad state at this excitation energy.

\subsubsection{State 37: 11.266 MeV}

A potential new state is observed at $E_x = 11.266$ MeV. However, this state is only observed at one angle and corresponds to no known state in $^{25}$Mg$+n$ experiments. If the state is genuine, it must have a small neutron width to have been missed in $^{25}$Mg$+n$ experiments \cite{Massimi20171,PhysRevC.85.044615}. No matching state in $^{24}$Mg exists.

\subsubsection{States 38-42}

These states are covered by the contaminating $^{16}$O peaks in the present data.

\subsubsection{Additional note concerning state 41: 11.289 MeV}

This state is not observed in the present experiment as it is covered by the contaminating $^{16}$O states. However, based on the observation of a state at 11.29(3) MeV in $^{26}$Mg($\alpha,\alpha^\prime$)$^{26}$Mg reactions with $J>1$ \cite{PhysRevC.96.055802}, which cannot be the $J^\pi = 2^-$ state (state 40 in the present work) at 11.295 MeV \cite{Massimi20171}, we conclude that there is a natural-parity state with $J>1$ at $E_x = 11.289$ MeV taking the energy of the state from from Ref. \cite{Massimi20171} and the assignment of the spin and parity from Ref. \cite{PhysRevC.96.055802}.

\subsubsection{States 43 and 44: The 11.32-11.33 MeV region}

There are two outstanding questions in this region: firstly, whether the lowest observed resonance at $E_\alpha^{\mathrm{lab}} = 832(2)$ keV ($E_x = 11.319(2)$ MeV) in $^{22}$Ne($\alpha,n$)$^{25}$Mg \cite{PhysRevLett.87.202501} corresponds to the resonance observed at $E_\alpha^{\mathrm{lab}} = 828(5)$ keV ($E_x = 11.315(5)$ MeV) in $^{22}$Ne($\alpha,\gamma$)$^{26}$Mg \cite{Wolke1989}, and the possible correspondence of this state or these states with the resonance observed in $^{25}$Mg$+n$ reactions at $E_n^{\mathrm{lab}} = 243.98(2)$ keV ($E_x = 11.328$ MeV) \cite{Massimi20171}.

In the present data, there is a state located at 11.321(1) MeV (state 43) and an additional state (number 44) located at 11.329(1) MeV. This second state is likely to be the state observed in $^{25}$Mg$+n$ reactions, a state which has not been observed in direct $^{22}$Ne($\alpha,n$)$^{25}$Mg measurements. We therefore conclude that the $E_x = 11.328$ MeV state observed in Ref. \cite{Massimi20171} is distinct from the resonance or resonances observed in Refs. \cite{Wolke1989,PhysRevLett.87.202501}. The $E_x = 11.328$ MeV state may have unnatural parity as suggested in Ref. \cite{Massimi20171}. A $J^\pi = 1^+$ state is known to exist at $E_x = 11.32(1)$ MeV \cite{PhysRevC.39.311} and we would tentatively make the connection between that state and the $E_x = 11.328$ MeV state of Ref. \cite{Massimi20171}.

We accept that one problem with our conclusion that the $E_n^{\mathrm{lab}} = 243.98(2)$ keV resonance in $^{25}$Mg$+n$ reactions is distinct from the $^{22}$Ne($\alpha,n$)$^{25}$Mg resonance is that the width of the resonance measured in Ref. \cite{PhysRevLett.87.202501} is inconsistent with the lack of an observed state in $^{25}$Mg$+n$ reactions as otherwise the $^{22}$Ne($\alpha,n$)$^{25}$Mg resonance would have been observed in Refs. \cite{Massimi20171,PhysRevC.85.044615}. Presently this problem is not resolved. Future experimental studies of the $^{22}$Ne($\alpha,n$)$^{25}$Mg reaction are required to resolve this discrepancy.

We note that, due to the close proximity of the $^{16}$O contamination it is not possible to reject the existence of a state at $E_x = 11.315$ MeV corresponding to the $^{22}$Ne($\alpha,\gamma$)$^{26}$Mg resonance of Ref. \cite{Wolke1989}. As such, we are not able to determine if the $^{22}$Ne($\alpha,\gamma$)$^{26}$Mg resonance of Ref. \cite{Wolke1989} and the $^{22}$Ne($\alpha,n$)$^{25}$Mg resonance of Ref. \cite{PhysRevLett.87.202501} correspond to the same state in $^{26}$Mg.

\subsubsection{State 45: 11.345 MeV}

Two levels have been observed in $^{25}$Mg($n,\gamma$)$^{26}$Mg at $E_x = 11.345$ MeV, one narrower ($\Gamma = 300-3900$ eV) and the second broader ($\Gamma = 6-9$ keV). In the present experiment, only one state is observed. This may be because the states are not resolved. Accordingly, we are unable to help to provide further limitations for the widths than already present in Refs. \cite{PhysRevC.85.044615,Massimi20171}.

\subsubsection{State 50: 11.414 MeV}

A potential new state is observed at $E_x = 11.414$ MeV. However, this state may correspond to the state in $^{24}$Mg at $E_x = 11.389$ MeV. If the state is real then the neutron width for the state must be small to have escaped detection in previous $^{25}$Mg$+n$ experiments \cite{PhysRevC.85.044615,Massimi20171}

\subsubsection{State 51: 11.426 MeV}

A potential new state is observed at $E_x = 11.426$ MeV. However, this state is only observed at one angle and corresponds to no known state in $^{25}$Mg$+n$ experiments. It may, however, correspond to a known state in $^{24}$Mg at $E_x = 11.453$ MeV. If the state is genuine, it must have a small neutron width to have been missed in $^{25}$Mg$+n$ experiments \cite{Massimi20171,PhysRevC.85.044615}.

\subsubsection{States 52: 11.444 MeV}

A state at $E_x = 11.444(1)$ MeV is observed in the present experiment. This state is assigned as $J^\pi = 4^+$ in Ref. \cite{PhysRevC.66.055805} by considering the heights of the peaks in the total cross section.

The measured resonance strength for the corresponding resonance is $\omega\gamma_{(\alpha,n)} = 0.034(4)$ meV. Under the assumption that the total width is dominated by the neutron width ($\Gamma \approx \Gamma_n$), the resonance strength is related to the $\alpha$-particle width by:

\begin{equation}
 \omega\gamma = (2J+1)\Gamma_\alpha.
 \end{equation}

This gives $\Gamma_\alpha = 3.7(4) \mu$eV assuming $J=4$. 

For a $J^\pi = 4^+$ state formed in $^{22}$Ne$+\alpha$ reactions, the $\alpha$ particle must have orbital angular momentum $\ell_\alpha = 4$. The single-particle limit for an $\ell_\alpha = 4$ $\alpha$-particle decay may be calculated \cite{RevModPhys.30.257} and is found to be $13.7$ $\mu$eV. The measured $\omega\gamma_{(\alpha,n)}$ therefore exhausts $27(3)\%$ of the single-particle strength.

While this is possible, one would expect that observed cross sections in $^{22}$Ne($^6$Li,$d$)$^{26}$Mg $\alpha$-cluster transfer reactions \cite{GIESEN199395} to be much greater for such a significant cluster state. In contrast, the measured $^{22}$Ne($^6$Li,$d$)$^{26}$Mg $\alpha$ cross section is more consistent with a spectroscopic factor of the order of a few percent.

For this reason, we suggest that the $J^\pi = 4^+$ assignment for this state is, at the very least, problematic and in need of further confirmation.


\subsubsection{State 54: 11.467 MeV}

It is useful to begin by discussing the various observations of states at around $E_x = 11.467$ MeV in $^{26}$Mg. In the present experiment, a state is observed at $E_x = 11.467(1)$ MeV with a width of $\Gamma = 6.2(4)$ keV.

A resonance at $E_\alpha^{\mathrm{lab}} = 1000$ keV ($E_x = 11.461(2)$ MeV) has been observed in $^{22}$Ne($\alpha,n$)$^{25}$Mg reactions. As this resonance has been observed in $^{22}$Ne($\alpha,n$)$^{25}$Mg reactions, it must have natural parity.

A state has also been observed at $E_n^{\mathrm{lab}} = 387.57$ keV ($E_x = 11.466$ MeV) using $^{25}$Mg$+n$ reactions, this state has a width of $\Gamma = 6.5 - 8.9$ keV depending on the source \cite{PhysRevC.85.044615,PhysRevC.66.055805}. Based on the height of the peak in $^{25}$Mg$+n$ data, Koehler \cite{PhysRevC.66.055805} assigns this state to have $J = 5$, and connects it to the resonance seen in $^{22}$Ne($\alpha,n$)$^{25}$Mg reactions. For this reason, a $J^\pi = 5^-$ assignment is made which has thereafter been used for computation of the $^{22}$Ne$+\alpha$ reaction rates \cite{PhysRevC.85.065809}.

A $J^\pi = 1^+$ state has been observed at $E_x = 11.46(1)$ MeV in $^{26}$Mg($p,p^\prime$)$^{26}$Mg reactions \cite{PhysRevC.39.311}. The $J^\pi = 1^+$ state cannot be the state observed in $^{22}$Ne($\alpha,n$)$^{25}$Mg reactions as it has unnatural parity. This state has been added to Table \ref{tab:states} for completeness.

In the case of a $J^\pi = 5^-$ assignment, as suggested in Ref. \cite{PhysRevC.66.055805}, the orbital angular momentum of the in-going $\alpha$ particle must be $\ell_\alpha = 5$. The single-particle limit for this $\alpha$-particle decay is 0.994 $\mu$eV \cite{RevModPhys.30.257}. The same logic applies as for the $11.444$-MeV state (state number 52), that the total width is dominated by the neutron width, and the resonance strength is given by $\omega\gamma = (2J+1)\Gamma_\alpha$. In the direct $^{22}$Ne($\alpha,n$)$^{25}$Mg measurement of Ref. \cite{PhysRevLett.87.202501}, the resonance strength is $\omega\gamma = 0.048(10)$ meV which is $4.4$ times greater than the single-particle limit. The cross section measured in the $^{22}$Ne($^6$Li,$d$)$^{26}$Mg reaction is again more consistent with a spectroscopic factor of a few percent of the single-particle limit. This suggests that either the assignment of $\ell_\alpha = 5$ for this resonance is incorrect or that the directly measured resonance strength is too high.

Additionally, a $J^\pi = 5^-$ resonance would require a neutron orbital momentum of $\ell_n = 3$ to be populated from the $J^\pi = 5/2^+$ ground state of $^{25}$Mg. Computing the single-particle limit for this $\ell_n = 3$ decay results in a limit of $0.75$ keV, which is about an order-of-magnitude smaller than the measured widths which are in the range of $\Gamma = 6.5 - 9.3$ keV \cite{PhysRevC.85.044615,PhysRevC.66.055805}. As the R-matrix analyses in Refs. \cite{Massimi20171,PhysRevC.85.044615,PhysRevC.66.055805} do not include contributions from $\ell_n>2$, these analyses would not have been able to exclude an $\ell_n=3$ assignment on the basis of the width of the state. 

It is not clear whether the level observed in $^{25}$Mg$+n$ reactions is the $1^+$ state observed in the $^{26}$Mg($p,p^\prime$)$^{26}$Mg reaction \cite{PhysRevC.39.311} or the state observed in the $^{22}$Ne($^6$Li,$d$)$^{26}$Mg reaction and the $^{22}$Ne($\alpha,n$)$^{25}$Mg reaction. It is also possible that both levels could have been observed but incorrectly treated as one level in Ref. \cite{PhysRevC.66.055805}. A reevalulation of the $^{25}$Mg$+n$ data at higher incident neutron energies with $R$-matrix analysis including higher-$\ell$ partial waves may help to clarify the properties of the levels at this excitation energy.

\subsubsection{State 55: 11.481 MeV}

A potential new state has been observed at $11.481(1)$ MeV. However, this state may correspond to the state in $^{24}$Mg at $E_x = 11.456$ MeV. If the state is not a contaminant, then it must have a small neutron width to have escaped detection in $^{25}$Mg$+n$ reactions \cite{PhysRevC.85.044615,Massimi20171}

\section{Conclusions and Outlook}

Excited states of $^{26}$Mg were studied in high resolution using the Q3D spectrograph at MLL, Garching. Clarification of the number and location of states resolving some of the discrepancies noted in Ref. \cite{PhysRevC.96.055802} was given, notably the observation of multiple levels just above 10.8 MeV. Four new levels (states 16, 22, 24 and 26 at $10.943$, $11.074$, $11.102$ and $11.119$ MeV) were definitively observed in $^{26}$Mg. The $11.102$- and $11.119$-MeV states are above the neutron threshold but were not observed in $^{25}$Mg$+n$ reactions implying that these states have small neutron widths. It is unknown whether these levels contribute to $\alpha$-particle-induced reactions on $^{22}$Ne as no information on the $J^\pi$ of these states is available.

Up to six additional potential levels (states 33, 34, 37, 50, 51 and 55 at $11.209$, $11.216$, $11.266$, $11.414$, $11.426$ and $11.481$ MeV) were observed in $^{26}$Mg but these cannot yet be confirmed. Some of the potential new levels could be due to $^{24}$Mg contamination in the target. All of these potential levels are above the neutron threshold.

One of the previously observed natural-parity levels above the neutron threshold  in $^{26}$Mg ($E_x = 11.113$ MeV with $J^\pi = 2^+$) is populated extremely weakly in the $^{26}$Mg($d,d^\prime$)$^{26}$Mg reaction suggesting that the state may have isospin $T=2$ and a correspondingly small contribution to the $^{22}$Ne$+\alpha$ reaction rates.

A level (43) is observed at $11.321$ MeV probably corresponding to the $E_\alpha^{\mathrm{lab}} =  832$-keV resonance observed in $^{22}$Ne($\alpha,n$)$^{25}$Mg reactions \cite{PhysRevLett.87.202501}. Another level (44) is observed at $11.329$ MeV probably corresponding to the $E_n^{\mathrm{lab}} = 243.98$-keV resonance observed in $^{25}$Mg$+n$ reactions. This suggests that the width of the resonance in $^{22}$Ne($\alpha,n$)$^{25}$Mg may have been over-estimated. A remeasurement of this level is probably required to solve the inconsistency in the available nuclear data.

The spins and resonance strengths of the $E_x = 11.426$- and $11.467$-MeV states (numbers  also need to be verified as the present nuclear data are inconsistent. The spin assignments of the levels could be incorrect, the resonance strengths overestimated or the levels observed in $^{25}$Mg$+n$ reactions may not be the same as the levels observed in $^{22}$Ne($\alpha,n$)$^{25}$Mg reactions.

There are now obvious avenues in studying the structure of $^{26}$Mg. In particular, future experimental studies of the astrophysically important resonances in $^{26}$Mg can try to compare observed states with the states observed in the present study. The spins and parities of those states without assignments need to be determined so that a list of the states which may contribute to the $^{22}$Ne$+\alpha$ reactions can be compiled, and estimates for the $\alpha$-particle partial widths of these states need to be made.

Future direct measurements which are able to verify the total widths of some of the higher-energy states would also be beneficial. This may help to resolve some of the outstanding questions as to which states observed in $^{25}$Mg$+n$ reactions correspond to known $^{22}$Ne($\alpha,n$)$^{25}$Mg resonances, and may therefore help with the associated spin assignments for these states, and lead in due course to a re\"{e}valuation of the astrophysical reaction rates.

\section{Acknowledgements}

The authors wish to thank the beam operators at MLL for the stable high-quality beams delivered. RN acknowledges financial support from the NRF through grant number 85509. PA thanks Gavin Lotay and Richard Longland for useful discussions regarding $^{26}$Mg level assignments.

\bibliography{Mg26_MunichQ3D}

\end{document}